%% file: v2_Vaibhav.tex
\documentclass[conference]{IEEEtran}
\PassOptionsToPackage{ruled, linesnumbered}{algorithm2e}
\usepackage[latin9]{inputenc}
\usepackage{color}
\usepackage{verbatim}
\usepackage{mathrsfs}
\usepackage{algorithm2e}
\usepackage{amsmath}
\usepackage{amsthm}
\usepackage{amssymb}
\usepackage{graphicx}

\makeatletter
\theoremstyle{plain}
\newtheorem{thm}{\protect\theoremname}
\theoremstyle{remark}
\newtheorem{rem}[thm]{\protect\remarkname}

\usepackage{empheq}

\usepackage{balance}
\allowdisplaybreaks
\usepackage{url}
\usepackage{pgfplots}
\usepackage{pgfplotstable}
\usepackage{tikz}
\usetikzlibrary{calc}
\usepackage{cite}
\usepackage{xcolor}
\usepackage{mathtools}
\definecolor{myblue}{rgb}{0.0, 0.5, 1.0}
\definecolor{myred}{rgb}{1.0, 0.13, 0.32}
\definecolor{mygreen}{rgb}{0.31, 0.78, 0.47}

\newcommand{\herm}{^{\mathsf{H}}}
\newcommand{\trans}{^{\mathsf{T}}}

\DeclareMathOperator{\minimize}{minimize}
\DeclareMathOperator{\diag}{\mathsf{diag}}
\DeclareMathOperator{\maximize}{maximize}
\DeclareMathOperator{\st}{subject~to}

\makeatother

\providecommand{\remarkname}{Remark}
\providecommand{\theoremname}{Theorem}

\begin{document}
\title{{\huge{SCA-Based Beamforming Optimization for IRS-Enabled Secure
Integrated Sensing and Communication}}}
\author{Vaibhav Kumar\IEEEauthorrefmark{1}, Marwa Chafii\IEEEauthorrefmark{2}\IEEEauthorrefmark{3},
A. Lee Swindlehurst\IEEEauthorrefmark{4}, Le-Nam Tran\IEEEauthorrefmark{1},
and Mark F. Flanagan\IEEEauthorrefmark{1}\\
\IEEEauthorblockA{\IEEEauthorrefmark{1}School of Electrical and Electronic Engineering,
University College Dublin, Belfield, Dublin 4, Ireland\\
\IEEEauthorrefmark{2}Engineering Division, New York University (NYU),
Abu Dhabi, UAE\\
\IEEEauthorrefmark{3}NYU WIRELESS, NYU Tandon School of Engineering,
New York, USA\\
\IEEEauthorrefmark{3}Center for Pervasive Communications and Computing,
University of California, Irvine, CA 92697 USA\\
Email: vaibhav.kumar@ieee.org, marwa.chafii@nyu.edu, swindle@uci.edu,
\\
nam.tran@ucd.ie, mark.flanagan@ieee.org}}
\maketitle
{\let\thefootnote\relax\footnotetext{This work was supported by the Irish Research Council under Grant IRCLA/2017/209, and also in part by Science Foundation Ireland under Grant 17/CDA/4786.}}
\begin{abstract}
Integrated sensing and communication (ISAC) is expected to be offered
as a fundamental service in the upcoming sixth-generation (6G) communications
standard. However, due to the exposure of information-bearing signals
to the sensing targets, ISAC poses unique security challenges. In
recent years, intelligent reflecting surfaces (IRSs) have emerged
as a novel hardware technology capable of enhancing the physical layer
security of wireless communication systems. Therefore, in this paper,
we consider the problem of transmit and reflective beamforming design
in a secure IRS-enabled ISAC system to maximize the beampattern gain
at the target. The formulated non-convex optimization problem is challenging
to solve due to the intricate coupling between the design variables.
Moreover, alternating optimization (AO) based methods are inefficient
in finding a solution in such scenarios, and convergence to a stationary
point is not theoretically guaranteed. Therefore, we propose a novel
successive convex approximation (SCA)-based second-order cone programming
(SOCP) scheme in which all of the design variables are updated simultaneously
in each iteration. The proposed SCA-based method significantly outperforms
a penalty-based benchmark scheme previously proposed in this context.
Moreover, we also present a detailed complexity analysis of the proposed
scheme, and show that despite having slightly higher per-iteration
complexity than the benchmark approach the average problem-solving
time of the proposed method is notably lower than that of the benchmark
scheme.
\end{abstract}


\section{Introduction}

The sixth-generation (6G) wireless standard is being developed not
only to improve the quality of user experience compared to that offered
by the fifth-generation networks, but also to support a range of new
wireless communication services, such as autonomous vehicles, drone
monitoring, human activity recognition, environmental monitoring,
enhanced localization and tracking, and many more. Supporting such
new services will require the integration of communication, sensing
and localization capabilities as fundamental services in a single
network architecture rather than as auxiliary functionalities~\cite{22-ComMag-MultiFunc6g}.
Integrated sensing and communication (ISAC) has recently emerged as
a potential enabler in this direction, combining the communication
and sensing capabilities in a single hardware platform using a common
waveform~\cite{20-SPMag-AutoVeh,23-IoTJ-JCAS}. Preliminary results
have confirmed that ISAC can improve the spectral efficiency of a
network by virtue of exploiting a common hardware, signal processing
and spectral framework, thereby offering a low-cost solution to the
spectrum scarcity problem. Furthermore, by exploiting the possibility
of communication-centric and sensing-centric designs, it can also
enjoy significant coordination gains compared to conventional networks.
However, due to the broadcast nature of wireless channels and the
inclusion of information-bearing signaling in the sensing waveform,
susceptibility to eavesdropping targets poses unique security challenges
in ISAC.

Intelligent reflecting surfaces (IRSs) have recently emerged as a
groundbreaking hardware technology to robustify wireless communication
systems against eavesdroppers via passive beamforming~\cite{23-ComMag-IRS}.
The benefits of IRS in a secure communication-only multiple-input
multiple-output (MIMO) system have been well established in the literature~\cite{21-MILCOM-IRS-MIMOME,22-VTC-maxEE-MIMOME-IRS}.
Hence, it is worth exploring the advantages of IRSs in an ISAC system
in terms of physical layer security~\cite{22-VTMag-IRS-sISAC}. However,
it is interesting to note that in contrast to the somewhat rich literature
on IRS-aided ISAC systems~\cite{23-WC-ISAC-RIS-Mag,22-WCNC-IRS-ISAC,22-TVT-QLiu-RIS-ISAC,22-EUSIPCO-Swindle,22-EUSIPCO-HybridRIS},
there is a dearth of literature on \textit{secure} IRS-aided ISAC
system design~\cite{23-TVT-activeRIS-sISAC,22-arXiv-sISAC}. As one
of the few examples, the authors in~\cite{23-TVT-activeRIS-sISAC}
considered an active IRS-aided multiuser multiple-input single-output
(MU-MISO) ISAC system, where the aim was to obtain an optimal beamforming
design that maximizes the achievable secrecy rate of the communication
users while guaranteeing a minimum radar signal-to-interference-plus-noise
ratio (SINR). In~\cite{22-arXiv-sISAC}, the authors considered
the problem of beampattern optimization for an eavesdropping target
in an IRS-enabled MU-MISO secure ISAC system, subject to SINR constraints
at the communication users and information leakage constraints at
the target. Two different scenarios were considered in~\cite{22-arXiv-sISAC};
in the first scenario, full channel state information (CSI) and the
target location were assumed to be known at the base station (BS),
while imperfect CSI and uncertain target location were assumed in
the second scenario. In this paper, we will focus on the first scenario
only, where the CSI and target location are known at the BS. The more
practical setting in which these quantities are imprecisely known
will be considered in future work. To optimize the transmit and IRS
beamforming in the first scenario, the authors in~\cite{22-arXiv-sISAC}
proposed a penalty-based alternating optimization (AO) algorithm to
obtain a semi-closed-form solution using Lagrange duality and a majorization-minimization
(MM) algorithm.

\begin{figure}[t]
\begin{centering}
\includegraphics[width=0.80\columnwidth]{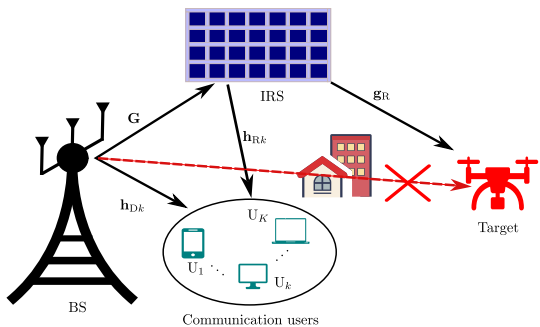}
\par\end{centering}
\caption{System model for IRS-enabled secure ISAC system.}
\label{fig:SysMod}
\end{figure}

Even though the use of AO in~\cite{22-arXiv-sISAC} makes the optimization
problem much easier to solve, it may not produce a high-quality solution
because of the complicated interdependence between the design variables~\cite{21-WCmag-optTech}.
Moreover, as we will show in~Section~\ref{sec:Results}, the use
of a penalty-based method requires a large number of iterations to
achieve convergence and therefore has a very high problem-solving
time. Note also that a feasible solution is not guaranteed if the
algorithm terminates prematurely. To tackle these issues, in this
paper we propose a successive convex approximation (SCA) based beampattern
optimization scheme which results in a high-performance solution and
also requires a much shorter convergence time. The main contributions
of the paper are listed as follows:
\begin{itemize}
\item We propose a provably convergent SCA-based algorithm to maximize the
beampattern gain at the eavesdropping target in the secure ISAC system,
subject to the SINR requirements at the communication users and information
leakage constraints at the target. In contrast to the AO-based scheme
of~\cite{22-arXiv-sISAC} where all design variables are updated
in an alternating fashion, we derive a second-order cone program (SOCP)
where all of the optimization variables are updated \textit{simultaneously}
in each iteration.
\item We present a complexity analysis of the proposed scheme which demonstrates
that the per-iteration complexity grows as $\mathcal{O}\big(N^{3.5}\big)$,
while that of the benchmark solution is given by $\mathcal{O}\big(N^{3}\big)$~\cite[Sec.~III-C]{22-arXiv-sISAC},
where $N$ is the number of reflecting elements in the IRS. Although
the per-iteration complexity of the proposed approach is slightly
higher than that of the benchmark scheme, we show that it requires
significantly fewer iterations to converge, resulting in a much shorter
problem-solving time.
\item We present extensive numerical results to confirm that the proposed
SCA-based SOCP approach results in a high-performance solution, and
significantly outperforms the penalty-based AO algorithm in~\cite{22-arXiv-sISAC}.
\end{itemize}

\paragraph*{Notation}

Bold uppercase and lowercase letters are used to denote matrices and
vectors, respectively. By $\mathbb{C}^{M\times N}$, we denote the
vector space of all $M\times N$ complex-valued matrices. By $\mathbf{X}\trans$,
$\mathbf{X}\herm$, $\|\mathbf{X}\|$, $\Re\{\mathbf{X}\}$ and $\Im\{\mathbf{X}\}$,
we respectively denote the transpose, conjugate transpose, Frobenius
norm, real and imaginary components of a matrix $\mathbf{X}$. $|x|$
denotes the absolute value of a complex number $x$, and $\diag(x)$
denotes the diagonal matrix whose main diagonal comprises the elements
of $\mathbf{x}$. $\mathcal{O}(\cdot)$ denotes the Bachmann--Landau
notation.

\section{System Model and Problem Formulation}

Consider the ISAC system shown in Fig.~\ref{fig:SysMod} consisting
of a multi-antenna dual-function base station (BS), one IRS, $K$
single-antenna communication users (denoted by $\text{U}_{k},k\in\mathcal{K}\triangleq\{1,2,\ldots,K\}$),
and one single-antenna eavesdropping target.\footnote{Although we consider a single eavesdropping target in this paper,
it is straightforward to use the proposed algorithm for a system with
multiple eavesdropping targets.} Let $L$ and $N$ denote the number of antennas at the BS and the
number of elements in the IRS, respectively. We assume that the BS
transmits a linear superposition of radar and information signals
for the purpose of joint sensing and communication. The signal vector
transmitted from the BS is given by 
\begin{equation}
\mathbf{s}=\sum\nolimits _{k\in\mathcal{K}}\mathbf{x}_{k}w_{k}+\sum\nolimits _{l\in\mathcal{L}}\hat{\mathbf{x}}_{l}\hat{w}_{l},\label{eq:txSignal}
\end{equation}
where $w_{k}$ is the communication signal intended for $\text{U}_{k}$
and $\hat{w}_{l}$ is the $l^{\text{th}}$ radar signal with $l\in\mathcal{L}\triangleq\{1,2,\ldots,L\}$.
Moreover, $\mathbf{x}_{k}\in\mathbb{C}^{L\times1}$ and $\hat{\mathbf{x}}_{l}\in\mathbb{C}^{L\times1}$
are the beamforming vectors corresponding to $w_{k}$ and $\hat{w}_{l}$,
respectively. \textcolor{black}{It is assumed that $\mathbb{E}\big\{ w_{k}\big\}=0$,
$\mathbb{E}\big\{|w_{k}|^{2}\big\}=1\ \forall k\in\mathcal{K}$, $\mathbb{E}\big\{\hat{w}_{l}\big\}=0$,
$\mathbb{E}\big\{|\hat{w}_{l}|^{2}\big\}=1\ l\in\mathcal{L}$, and
$\mathbb{E}\big\{ w_{k}\hat{w}_{l}\herm\big\}=0\ \forall k\in\mathcal{K},l\in\mathcal{L}$,
i.e., the communication and radar signals are mutually independent
and uncorrelated.}\textcolor{blue}{}%
{} Denoting the BS-IRS, BS-$\text{U}_{k}$ and IRS-$\text{U}_{k}$ links
by $\mathbf{G}\in\mathbb{C}^{N\times L}$, $\mathbf{h}_{\text{D}k}\in\mathbb{C}^{1\times L}$,
and $\mathbf{h}_{\text{R}k}\in\mathbb{C}^{1\times N}$, respectively,
the signal received at $\text{U}_{k}$ is given by 
\begin{equation}
y_{k}=\mathbf{h}_{k}\mathbf{s}+z_{k},\label{eq:rxSignal-Uk}
\end{equation}
where $\mathbf{h}_{k}\triangleq\mathbf{h}_{\text{D}k}+\mathbf{h}_{\text{R}k}\boldsymbol{\Theta}\mathbf{G}$,
$\boldsymbol{\Theta}\triangleq\diag(\boldsymbol{\theta})$, $\boldsymbol{\theta}=[\theta_{1},\theta_{2},\ldots,\theta_{N}]\trans$,
$\theta_{n}\triangleq\exp(j2\pi\phi_{n})$ with $\phi_{n}\in[0,2\pi)$
denoting the phase shift induced by the $n^{\text{th}}$ IRS element,
and $z_{k}$ denotes the zero-mean complex additive white Gaussian
noise (AWGN) at $\text{U}_{k}$ with variance $\sigma_{k}^{2}$ .

Defining $\mathbf{X}\triangleq[\mathbf{x}_{1},\mathbf{x}_{2},\ldots,\mathbf{x}_{K},\hat{\mathbf{x}}_{1},\hat{\mathbf{x}}_{2},\ldots\hat{\mathbf{x}}_{L}]\in\mathbb{C}^{L\times(K+L)}$,
and $\tilde{\mathbf{x}}_{m}$ as the $m^{\text{th}}$ column of $\mathbf{X}$,
the signal-to-interference-plus-noise ratio (SINR) at $\text{U}_{k}$
to decode the intended message is given by 
\begin{equation}
\gamma_{k}=\frac{|\mathbf{h}_{k}\mathbf{x}_{k}|^{2}}{\sigma_{k}^{2}+\sum_{\ell\in\mathcal{M}\setminus\{k\}}|\mathbf{h}_{k}\tilde{\mathbf{x}}_{\ell}|^{2}},\label{eq:defSINR-Uk}
\end{equation}
where $\mathcal{M}\triangleq\{1,2,\ldots,K,K+1,\ldots,K+L\}.$ Similarly,
by denoting the IRS-target channel by $\mathbf{g}_{\text{R}}\in\mathbb{C}^{1\times N}$
and assuming that the BS-target (direct) link is blocked due to obstacles,
the signal received at the target is given by 
\begin{equation}
y_{\text{T}}=\mathbf{gs}+z_{\text{T}},\label{eq:rxSignal-T}
\end{equation}
where $\mathbf{g}\triangleq\mathbf{g}_{\text{R}}\boldsymbol{\Theta}\mathbf{G}\in\mathbb{C}^{1\times L}$,
\textcolor{black}{$\mathbf{g}_{\mathrm{R}}$ is the steering vector
from the IRS in the direction of the target, }and $z_{\text{T}}$
is the zero-mean complex AWGN at the target with variance $\sigma_{\text{T}}^{2}$.
Therefore, the SINR at the target to wiretap the signal intended for
$\text{U}_{k}$ is given by 
\begin{equation}
\hat{\gamma}_{k}=\frac{|\mathbf{g}\mathbf{x}_{k}|^{2}}{\sigma_{\text{T}}^{2}+\sum\nolimits _{\ell\in\mathcal{M}\setminus\{k\}}|\mathbf{g}\tilde{\mathbf{x}}_{\ell}|^{2}}.\label{eq:defSINR-Tk}
\end{equation}
We assume that all of the channels and the target location are perfectly
known to the BS. The beampattern gain toward the target is then given
by (c.f.~\cite{22-arXiv-sISAC})
\begin{equation}
\mathscr{G}(\mathbf{X},\boldsymbol{\theta})=\mathbb{E}\{|\mathbf{gs}|^{2}\}=\sum_{m\in\mathcal{M}}|\mathbf{g}\tilde{\mathbf{x}}_{m}|^{2}.\label{eq:defBPG}
\end{equation}
Therefore, the problem of maximizing the beampattern gain toward the
target is given by 
\begin{subequations}
\label{eq:mainProb}
\begin{align}
\underset{\mathbf{X},\boldsymbol{\theta}}{\maximize}\  & \mathscr{G}(\mathbf{X},\boldsymbol{\theta}),\label{eq:mainObj}\\
\st\  & \gamma_{k}\geq\Gamma_{k}\ \forall k\in\mathcal{K},\label{eq:SINRC-Uk}\\
 & \hat{\gamma}_{k}\leq\hat{\Gamma}_{k}\ \forall k\in\mathcal{K},\label{eq:SINRC-Tk}\\
 & \|\mathbf{X}\|\leq\sqrt{P}\label{eq:TPC}\\
 & |\theta_{n}|^{2}=1\ \forall n\in\mathcal{N}\triangleq\{1,2,\ldots,N\},\label{eq:UMC}
\end{align}
\end{subequations}
where~\eqref{eq:SINRC-Uk} ensures that the SINR at $\text{U}_{k}$
is greater than or equal to the predefined threshold $\Gamma_{k}$,~\eqref{eq:SINRC-Tk}
enforces the constraint that the maximum leakage of the information
intended for $\text{U}_{k}$ at the eavesdropping target is below
the tolerance level $\hat{\Gamma}_{k}$ and $P$ is the transmit power
budget at the BS. Note that in a system with heterogeneous secrecy
requirements, considering information leakage constraints results
in a more flexible resource allocation compared to that offered via
imposing constraints on the achievable secrecy rate~\cite{20-JSAC-infoLeakModel}.
It is easy to note that due to coupling between the design variables
$\mathbf{X}$ and $\boldsymbol{\theta}$ in~\eqref{eq:mainObj}--\eqref{eq:SINRC-Tk}
and the non-convex constraints in~\eqref{eq:UMC}, the problem in~\eqref{eq:mainProb}
is non-convex and challenging to solve.

Hua \textit{et al.}~\cite{22-arXiv-sISAC} proposed a penalty-based
dual-loop AO algorithm to obtain a solution to~\eqref{eq:mainProb}.
More specifically, in the inner loop, auxiliary variables were updated
by solving a quadratically-constrained quadratic program (QCQP), the
beamformers (i.e., $\mathbf{X}$) were updated using a bisection search,
and the IRS reflection coefficients ($\boldsymbol{\theta}$) were
updated via MM; the outer loop was used to update the penalty parameter
only. Although the use of auxiliary variables and the penalty method
in~\cite{22-arXiv-sISAC} resulted in a reformulated optimization
problem where the design variables were decoupled in the constraints,
obtaining a high quality solution is not guaranteed via AO. \textcolor{black}{Additionally,
although the per-iteration complexity of the penalty-based solution
in~\cite{22-arXiv-sISAC} was $\mathcal{O}\big(N^{3}\big)$, beca}use
of the use of bisection search and the dual-loop structure, the number
of iterations required for convergence is large. This in turn results
in a high problem-solving time because if the iterations are terminated
prematurely (i.e., before the penalty terms becomes nearly zero),
the obtained solution may not be feasible.

\section{Proposed Solution}

In this section, we apply a series of convex approximations to tackle
the non-convexity of~\eqref{eq:mainProb} and to obtain a high-performance
solution. In this regard, for two arbitrary complex-valued vectors
$\mathbf{u}$ and $\mathbf{v}$, we recall the following (in)equalities
(c.f.~\cite[eqn. (6)]{23-WCL-SCA})
\begin{subequations}
\label{eq:baiscINEQ}
\begin{align}
\|\mathbf{u}\|^{2} & \ \geq2\Re\{\mathbf{v}\herm\mathbf{u}\}-\|\mathbf{v}\|^{2},\label{eq:Inequality}\\
\Re\{\mathbf{u}\herm\mathbf{v}\} & \ =\frac{1}{4}\big(\|\mathbf{u}+\mathbf{v}\|^{2}-\|\mathbf{u}-\mathbf{v}\|^{2}\big),\label{eq:EQ1}\\
\Im\{\mathbf{u}\herm\mathbf{v}\} & \ =\frac{1}{4}\big(\|\mathbf{u}-j\mathbf{v}\|^{2}-\|\mathbf{u}+j\mathbf{v}\|^{2}\big).\label{eq:EQ2}
\end{align}
\end{subequations}
Next, we note that the term in~\eqref{eq:mainObj} is neither convex
nor concave. Since we want to maximize the function in~\eqref{eq:mainObj},
we obtain a corresponding \textit{concave lower bound} as follows:
\begin{align}
 & \mathscr{G}(\mathbf{X},\boldsymbol{\theta})=\sum_{m\in\mathcal{M}}|\mathbf{g}\tilde{\mathbf{x}}_{m}|^{2}\nonumber \\
\overset{(\texttt{a})}{\geq} & \ \sum_{m\in\mathcal{M}}\big[2\Re\{a_{m}^{(i)}\!\!\ \herm\mathbf{g}\tilde{\mathbf{x}}_{m}\}-|a_{m}^{(i)}|^{2}\big]\nonumber \\
\overset{(\texttt{b})}{=} & \ \sum_{m\in\mathcal{M}}\Big[\frac{1}{2}\big\{\|a_{m}^{(i)}\mathbf{g}\herm+\tilde{\mathbf{x}}_{m}\|^{2}\!-\!\|a_{m}^{(i)}\mathbf{g}\herm\!-\tilde{\mathbf{x}}_{m}\|^{2}\big\}\!-\!|a_{m}^{(i)}|^{2}\Big]\nonumber \\
\overset{(\texttt{c})}{\geq} & \ \sum_{m\in\mathcal{M}}\Big[\Re\big\{\mathbf{b}_{m}^{(i)}\!\!\ \herm\big[a_{m}^{(i)}\mathbf{g}\herm+\tilde{\mathbf{x}}_{m}\big]\big\}-\frac{1}{2}\|\mathbf{b}_{m}^{(i)}\|^{2}\nonumber \\
 & \qquad\qquad-\frac{1}{2}\|a_{m}^{(i)}\mathbf{g}\herm-\tilde{\mathbf{x}}_{m}\|^{2}-|a_{m}^{(i)}|^{2}\Big]\nonumber \\
\triangleq & \ \sum_{m\in\mathcal{M}}f_{m}(\tilde{\mathbf{x}}_{m},\boldsymbol{\theta};\tilde{\mathbf{x}}_{m}^{(i)},\boldsymbol{\theta}^{(i)}),\label{eq:objLowerBound}
\end{align}
where $\tilde{\mathbf{x}}_{m}^{(i)}$ and $\boldsymbol{\theta}^{(i)}$
denote the value of $\tilde{\mathbf{x}}_{m}$ and $\boldsymbol{\theta}$
in the $i^{\text{th}}$ iteration of the SCA process, respectively.
Moreover, $(\texttt{a})$ and $(\texttt{c})$ follow from~\eqref{eq:Inequality},
and $(\texttt{b})$ follows from~\eqref{eq:EQ1}. Additionally in~\eqref{eq:objLowerBound},
$a_{m}^{(i)}\triangleq\mathbf{g}^{(i)}\tilde{\mathbf{x}}_{m}^{(i)}$,
$\mathbf{b}_{m}^{(i)}\triangleq a_{m}^{(i)}\mathbf{g}^{(i)}\!\!\ \herm+\tilde{\mathbf{x}}_{m}^{(i)}$,
and $\mathbf{g}^{(i)}\triangleq\mathbf{g}_{\text{R}}\boldsymbol{\Theta}^{(i)}\mathbf{G}$.
Note that $f_{m}(\tilde{\mathbf{x}}_{m},\boldsymbol{\theta};\tilde{\mathbf{x}}_{m}^{(i)},\boldsymbol{\theta}^{(i)})$
is \textit{jointly concave} with respect to (w.r.t.) $\tilde{\mathbf{x}}_{m}$
and $\boldsymbol{\theta}$.

Next, we turn our attention to the non-convex constraints in~\eqref{eq:SINRC-Uk}.
Using~\eqref{eq:defSINR-Uk}, for any $k\in\mathcal{K}$, we can
equivalently represent~\eqref{eq:SINRC-Uk} as follows:
\begin{subequations}
\label{eq:commSINR-1}
\begin{align}
\frac{1}{\Gamma_{k}}\big|\mathbf{h}_{k}\mathbf{x}_{k}\big|^{2} & \geq\sigma_{k}^{2}+\sum_{\ell\in\mathcal{M}\setminus\{k\}}\big(\wp_{k\ell}^{2}+\bar{\wp}_{k\ell}^{2}\big),\label{eq:cSINR-1-1}\\
\wp_{k\ell} & \geq\big|\Re\big\{\mathbf{h}_{k}\tilde{\mathbf{x}}_{\ell}\big\}\big|\ \forall\ell\in\mathcal{M}\setminus\{k\},\label{eq:cSINR-1-2}\\
\bar{\wp}_{k\ell} & \geq\big|\Im\big\{\mathbf{h}_{k}\tilde{\mathbf{x}}_{\ell}\big\}\big|\ \forall\ell\in\mathcal{M}\setminus\{k\}.\label{eq:cSINR-1-3}
\end{align}
\end{subequations}
It is easy to see that if~\eqref{eq:SINRC-Uk} is feasible, then
so is~\eqref{eq:commSINR-1} and vice versa. Note that the right-hand
side (RHS) of~\eqref{eq:cSINR-1-1} is convex, and we only need to
obtain a concave lower bound on the left-hand side (LHS) of~\eqref{eq:cSINR-1-1}.
Following a similar line of argument to~\eqref{eq:objLowerBound},
this can be done as follows:
\begin{align}
& \frac{1}{\Gamma_{k}}\big|\mathbf{h}_{k}\mathbf{x}_{k}\big|^{2} \geq\frac{1}{\Gamma_{k}}\Re\big\{\big(\mathbf{d}_{k}^{(i)}\!\!\ \herm\big)\big[c_{k}^{(i)}\mathbf{h}_{k}\herm+\mathbf{x}_{k}\big]\big\} -\frac{1}{2}\|\mathbf{d}_{k}^{(i)}\|^{2}\nonumber \\
 & \!\!\!-\frac{1}{2}\|c_{k}^{(i)}\mathbf{h}_{k}\herm-\mathbf{x}_{k}\|^{2}-\big|c_{k}^{(i)}\big|^{2} \triangleq\frac{1}{\Gamma_{k}}\bar{f}_{k}\big(\mathbf{x}_{k},\boldsymbol{\theta};\mathbf{x}_{k}^{(i)},\boldsymbol{\theta}^{(i)}\big), \!\label{eq:fBarDef}
\end{align}
where $c_{k}^{(i)}\triangleq\mathbf{h}_{k}^{(i)}\mathbf{x}_{k}^{(i)}$
and $\mathbf{d}_{k}^{(i)}\triangleq c_{k}^{(i)}\mathbf{h}_{k}^{(i)}\!\!\ \herm+\mathbf{x}_{k}^{(i)}$.

Using the fact that $u\geq|v|$ iff $u\geq v$ or $u\geq|-v|$, and
following~\eqref{eq:EQ1}, $\wp_{k\ell}$ in~\eqref{eq:cSINR-1-2}
can be equivalently written as 
\begin{subequations}
\label{eq:tkl-1}
\begin{align}
\wp_{k\ell} & \geq\Re\big\{\mathbf{h}_{k}\tilde{\mathbf{x}}_{\ell}\big\}=\frac{1}{4}\big(\|\mathbf{h}_{k}\herm+\tilde{\mathbf{x}}_{\ell}\|^{2}-\|\mathbf{h}_{k}\herm-\tilde{\mathbf{x}}_{\ell}\|^{2}\big),\label{eq:tkl-1-1}\\
\wp_{k\ell} & \geq-\Re\big\{\mathbf{h}_{k}\tilde{\mathbf{x}}_{\ell}\big\}=\frac{1}{4}\big(\|\mathbf{h}_{k}\herm-\tilde{\mathbf{x}}_{\ell}\|^{2}-\|\mathbf{h}_{k}\herm+\tilde{\mathbf{x}}_{\ell}\|^{2}\big).\label{eq:tkl-1-2}
\end{align}
\end{subequations}
Since the negative quadratic term in the RHS of~\eqref{eq:tkl-1-1}
results in its non-convexity, we use the inequality in~\eqref{eq:Inequality}
to convexify~\eqref{eq:tkl-1-1} as follows:
\begin{align}
\wp_{k\ell} & \geq\frac{1}{4}\big[\|\mathbf{h}_{k}\herm+\tilde{\mathbf{x}}_{\ell}\|^{2}-2\Re\big\{\big(\mathbf{h}_{k}^{(i)}-\tilde{\mathbf{x}}_{\ell}^{(i)}\!\!\ \herm\big)\big(\mathbf{h}_{k}\herm-\tilde{\mathbf{x}}_{\ell}\big)\big\}\nonumber \\
 & \quad+\|\mathbf{h}_{k}^{(i)}\!\!\ \herm-\tilde{\mathbf{x}}_{\ell}^{(i)}\|^{2}\big]\triangleq\mu_{k\ell}\big(\tilde{\mathbf{x}}_{\ell},\boldsymbol{\theta};\tilde{\mathbf{x}}_{\ell}^{(i)},\boldsymbol{\theta}^{(i)}\big).\label{eq:muklDef}
\end{align}
Following a similar argument,~\eqref{eq:tkl-1-2} yields
\begin{align}
\wp_{k\ell} & \geq\frac{1}{4}\big[\|\mathbf{h}_{k}\herm-\tilde{\mathbf{x}}_{\ell}\|^{2}-2\Re\big\{\big(\mathbf{h}_{k}^{(i)}+\tilde{\mathbf{x}}_{\ell}^{(i)}\!\!\ \herm\big)\big(\mathbf{h}_{k}\herm+\tilde{\mathbf{x}}_{\ell}\big)\big\}\nonumber \\
 & \quad+\|\mathbf{h}_{k}^{(i)}\!\!\ \herm+\tilde{\mathbf{x}}_{\ell}^{(i)}\|^{2}\big]\triangleq\bar{\mu}_{k\ell}\big(\tilde{\mathbf{x}}_{\ell},\boldsymbol{\theta};\tilde{\mathbf{x}}_{\ell}^{(i)},\boldsymbol{\theta}^{(n)}\big).\label{eq:muklHatDef}
\end{align}
Analogously,~\eqref{eq:cSINR-1-3} yields the following inequalities:

\begin{align}
\bar{\wp}_{k\ell} & \geq\frac{1}{4}\big[\|\mathbf{h}_{k}\herm-j\tilde{\mathbf{x}}_{\ell}\|^{2}-2\Re\big\{\big(\mathbf{h}_{k}^{(i)}-j\tilde{\mathbf{x}}_{\ell}^{(i)}\!\!\ \herm\big)\big(\mathbf{h}_{k}\herm+j\tilde{\mathbf{x}}_{\ell}\big)\big\}\nonumber \\
 & \quad+\|\mathbf{h}_{k}^{(i)}\!\!\ \herm+j\tilde{\mathbf{x}}_{\ell}^{(i)}\|^{2}\big]\triangleq\upsilon_{k\ell}\big(\tilde{\mathbf{x}}_{\ell},\boldsymbol{\theta};\tilde{\mathbf{x}}_{\ell}^{(i)},\boldsymbol{\theta}^{(i)}\big),\label{eq:uklDef}\\
\bar{\wp}_{k\ell} & \geq\frac{1}{4}\big[\|\mathbf{h}_{k}\herm+j\tilde{\mathbf{x}}_{\ell}\|^{2}-2\Re\big\{\big(\mathbf{h}_{k}^{(i)}+j\tilde{\mathbf{x}}_{\ell}^{(i)}\!\!\ \herm\big)\big(\mathbf{h}_{k}\herm-j\tilde{\mathbf{x}}_{\ell}\big)\big\}\nonumber \\
 & \quad+\|\mathbf{h}_{k}^{(i)}\!\!\ \herm-j\tilde{\mathbf{x}}_{\ell}^{(i)}\|^{2}\big]\triangleq\bar{\upsilon}_{k\ell}\big(\tilde{\mathbf{x}}_{\ell},\boldsymbol{\theta};\tilde{\mathbf{x}}_{\ell}^{(i)},\boldsymbol{\theta}^{(i)}\big).\label{eq:uklBarDef}
\end{align}

We now focus on the non-convex constraint in~\eqref{eq:SINRC-Tk},
which for any $k\in\mathcal{K}$, can be written as 
\begin{equation}
\hat{\gamma}_{k}\leq\hat{\Gamma}_{k}\Rightarrow\sigma_{\text{T}}^{2}+\sum_{\ell\in\mathcal{M}\setminus\{k\}}|\mathbf{g}\tilde{\mathbf{x}}_{\ell}|^{2}\geq\frac{1}{\hat{\Gamma}}|\mathbf{g}\mathbf{x}_{k}|^{2}.\label{eq:SINRC-Tk-1}
\end{equation}
Note that we need a \textit{concave lower bound} on the LHS of~\eqref{eq:SINRC-Tk-1},
and a \textit{convex upper bound} on the RHS. Similar to~\eqref{eq:objLowerBound},
the former can be obtained by linearizing the quadratic term in the
LHS as follows:

\begin{align}
 & \sigma_{\text{T}}^{2}+\sum_{\ell\in\mathcal{M}\setminus\{k\}}|\mathbf{g}\tilde{\mathbf{x}}_{\ell}|^{2} \geq \sigma_{\text{T}}^{2}+\sum_{\ell\in\mathcal{M}\setminus\{k\}}f_{\ell}(\tilde{\mathbf{x}}_{\ell},\boldsymbol{\theta};\tilde{\mathbf{x}}_{\ell}^{(i)},\boldsymbol{\theta}^{(i)}).\label{eq:SINRC-Tk-LHS}
\end{align}
On the other hand, a convex upper bound on $|\mathbf{g}\mathbf{x}_{k}|^{2}/\hat{\Gamma}$
is given by $(\tau_{k}^{2}+\bar{\tau}_{k}^{2})/\hat{\Gamma}$, where
$\tau_{k}\geq|\Re\{\mathbf{g}\mathbf{x}_{k}\}|$ and $\bar{\tau}_{k}\geq|\Im\{\mathbf{g}\mathbf{x}_{k}\}|$.
Therefore, using~\eqref{eq:SINRC-Tk-1},~\eqref{eq:SINRC-Tk-LHS}
and the preceding arguments, for a given $k\in\mathcal{K}$, the constraint
in~\eqref{eq:SINRC-Tk} can be equivalently written as 
\begin{subequations}
\label{eq:secureInequalities}
\begin{align}
\sigma_{\mathrm{T}}^{2}+\sum_{\ell\in\mathcal{M}\setminus\{k\}}f_{\ell}\big(\tilde{\mathbf{x}}_{\ell},\boldsymbol{\theta};\tilde{\mathbf{x}}_{\ell}^{(i)},\boldsymbol{\theta}^{(i)}\big) & \geq\frac{1}{\hat{\Gamma}_{k}}\big(\tau_{k}^{2}+\bar{\tau}_{k}^{2}\big),\label{eq:secureSINRC}\\
\tau_{k} & \geq\big|\Re\big\{\mathbf{g}\mathbf{x}_{k}\big\}\big|,\label{eq:secureSINRC-real}\\
\bar{\tau}_{k} & \geq\big|\Im\big\{\mathbf{g}\mathbf{x}_{k}\big\}\big|.\label{eq:secureSINRC-imag}
\end{align}
\end{subequations}
Again, it can be noted that if~\eqref{eq:SINRC-Tk} is feasible,
then so is~\eqref{eq:secureInequalities} and vice versa. Moreover,
following a similar set of arguments to those in~\eqref{eq:tkl-1}--\eqref{eq:uklBarDef},
lower bounds on $\tau_{k}$ and \textbf{$\bar{\tau}_{k}$ }in~\eqref{eq:secureSINRC-real}
and~\eqref{eq:secureSINRC-imag}, respectively, are given by
\begin{subequations}
\label{eq:etaChi}
\begin{align}
\tau_{k} & \geq\frac{1}{4}\big[\|\mathbf{g}\herm+\mathbf{x}_{k}\|^{2}-2\Re\big\{\big(\mathbf{g}^{(i)}-\mathbf{x}_{k}^{(i)}\!\!\ \herm\big)\big(\mathbf{g}\herm-\mathbf{x}_{k}\big)\big\}\nonumber \\
 & \quad+\|\mathbf{g}^{(i)}\!\!\ \herm-\mathbf{x}_{k}^{(i)}\|^{2}\big]\triangleq\eta_{k}\big(\mathbf{x}_{k},\boldsymbol{\theta};\mathbf{x}_{k}^{(i)},\boldsymbol{\theta}^{(i)}\big),\label{eq:etaDef}\\
\tau_{k} & \geq\frac{1}{4}\big[\|\mathbf{g}\herm-\mathbf{x}_{k}\|^{2}-2\Re\big\{\big(\mathbf{g}^{(i)}+\mathbf{x}_{k}^{(i)}\!\!\ \herm\big)\big(\mathbf{g}\herm+\mathbf{x}_{k}\big)\big\}\nonumber \\
 & \quad+\|\mathbf{g}^{(i)}\!\!\ \herm+\mathbf{x}_{k}^{(i)}\|^{2}\big]\triangleq\bar{\eta}_{k}\big(\mathbf{x}_{k},\boldsymbol{\theta};\mathbf{x}_{k}^{(i)},\boldsymbol{\theta}^{(i)}\big),\label{eq:etaBarDef}\\
\bar{\tau}_{k} & \geq\frac{1}{4}\big[\|\mathbf{g}\herm-j\mathbf{x}_{k}\|^{2}-2\Re\big\{\big(\mathbf{g}^{(i)}-j\mathbf{x}_{k}^{(i)}\!\!\ \herm\big)\big(\mathbf{g}\herm+j\mathbf{x}_{k}\big)\big\}\nonumber \\
 & \quad+\|\mathbf{g}^{(i)}\!\!\ \herm+j\mathbf{x}_{k}^{(i)}\|^{2}\big]\triangleq\chi_{k}\big(\mathbf{x}_{k},\boldsymbol{\theta};\mathbf{x}_{k}^{(i)},\boldsymbol{\theta}^{(i)}\big),\label{eq:chiDef}\\
\bar{\tau}_{k} & \geq\frac{1}{4}\big[\|\mathbf{g}\herm+j\mathbf{x}_{k}\|^{2}-2\Re\big\{\big(\mathbf{g}^{(i)}+j\mathbf{x}_{k}^{(i)}\!\!\ \herm\big)\big(\mathbf{g}\herm-j\mathbf{x}_{k}\big)\big\}\nonumber \\
 & \quad+\|\mathbf{g}^{(i)}\!\!\ \herm-j\mathbf{x}_{k}^{(i)}\|^{2}\big]\triangleq\bar{\chi}_{k}\big(\mathbf{x}_{k},\boldsymbol{\theta};\mathbf{x}_{k}^{(i)},\boldsymbol{\theta}^{(i)}\big).\label{eq:chiBarDef}
\end{align}
\end{subequations}

Next, since the constraint in~\eqref{eq:TPC} is already convex,
we are left only with the non-convexity of~\eqref{eq:UMC}. To tackle
this, we first relax the equality constraint in~\eqref{eq:UMC} by
a (convex) inequality constraint. In order to ensure that the inequality
constraint is satisfied with equality (i.e., the constraint is binding
at convergence), we add a regularization term in the objective and
handle the resulting non-convex objective by the first-order approximation
of the regularization term around $\boldsymbol{\theta}^{(i)}$. Therefore,
an equivalent reformulation of the problem in~\eqref{eq:mainProb}
can be given by 
\begin{subequations}
\label{eq:transProb-1}
\begin{align}
\underset{\mathbf{X},\boldsymbol{\theta},\boldsymbol{\wp},\bar{\boldsymbol{\wp}},\boldsymbol{\tau},\bar{\boldsymbol{\tau}}}{\maximize}\  & \sum_{m\in\mathcal{M}}f_{m}\big(\tilde{\mathbf{x}}_{m},\boldsymbol{\theta};\tilde{\mathbf{x}}_{m}^{(i)},\boldsymbol{\theta}^{(i)}\big)\nonumber \\
 & \qquad+\zeta\big[2\Re\big\{\boldsymbol{\theta}^{(n)}\ \herm\boldsymbol{\theta}\big\}-\|\boldsymbol{\theta}^{(n)}\|^{2}\big],\label{eq:transObj-1}\\
\st\  & \frac{1}{\Gamma_{k}}\bar{f}_{k}\big(\mathbf{x}_{k},\boldsymbol{\theta};\mathbf{x}_{k}^{(i)},\boldsymbol{\theta}^{(i)}\big)\nonumber \\
 & \quad\geq\sigma_{k}^{2}+\!\!\!\!\sum_{\ell\in\mathcal{M}\setminus\{k\}}\!\!\!\!\big(\wp_{k\ell}^{2}+\bar{\wp}_{k\ell}^{2}\big)\ \forall k\in\mathcal{K},\label{eq:transSINRC-1}\\
 & \eqref{eq:muklDef}-\eqref{eq:uklBarDef}\ \forall k\in\mathcal{K},\forall\ell\in\mathcal{M}\setminus\{k\},\label{eq:transSINRC-2}\\
 & \eqref{eq:secureSINRC},\eqref{eq:etaChi}\ \forall k\in\mathcal{K},\nonumber \\
 & \eqref{eq:TPC},\nonumber \\
 & |\theta_{n}|\leq1\ \forall n\in\mathcal{N},\label{eq:relaxedUMC}
\end{align}
\end{subequations}
where $\boldsymbol{\wp}\triangleq[\wp_{11},\wp_{12},\ldots,\wp_{KL}]\trans$,
$\bar{\boldsymbol{\wp}}\triangleq[\bar{\wp}_{11},\bar{\wp}_{12},\ldots\bar{\wp}_{KL}]\trans$,
$\boldsymbol{\tau}\triangleq[\tau_{1},\tau_{2},\ldots,\tau_{K}]\trans$,
$\bar{\boldsymbol{\tau}}\triangleq[\bar{\tau}_{1},\bar{\tau}_{2},\ldots,\bar{\tau}_{K}]\trans$,
and $\zeta>0$ is the regularization parameter. It is straightforward
to show that all of the constraints in~\eqref{eq:transProb-1} can
be represented by quadratic cones, and therefore~\eqref{eq:transProb-1}
is an SOCP problem which can be solved efficiently using off-the-shelf
solvers, e.g., MOSEK~\cite{mosek}. The proposed SCA-based SOCP method
is outlined in~\textbf{Algorithm~\ref{algo}}. 
\begin{algorithm}[t]
\caption{Proposed SCA-based Method to Solve~\eqref{eq:transProb-1}.}

\label{algo}

\KwIn{$\mathbf{X}^{(0)}$, $\boldsymbol{\theta}^{(0)}$, $\xi>0$}

$i\leftarrow0$\;

\Repeat{convergence }{

Solve~\eqref{eq:transProb-1} and denote the solution as $\mathbf{X}^{\star}$,
$\boldsymbol{\theta}^{\star}$\;

Update: $\mathbf{X}^{(i+1)}\leftarrow\mathbf{X}^{\star}$, $\boldsymbol{\theta}^{(i+1)}\leftarrow\boldsymbol{\theta}^{\star}$\;

$i\leftarrow i+1$\;

}

\KwOut{$\mathbf{X}^{\star}$, $\boldsymbol{\theta}^{\star}$}
\end{algorithm}

\begin{rem}
One needs to find feasible starting points $\mathbf{X}^{(0)}$ and
$\boldsymbol{\theta}^{(0)}$ to run ~\textbf{Algorithm~\ref{algo}},
which is not straightforward. Therefore, below we describe a practical
way to obtain a set of initial points. Consider the following optimization
problem: 
\begin{subequations}
\label{eq:initialPointProb}
\begin{align}
\underset{\mathbf{X},\boldsymbol{\theta},\boldsymbol{\delta},\bar{\boldsymbol{\delta}}}{\minimize}\  & \sum_{k\in\mathcal{K}}(\delta_{k}+\bar{\delta}_{k}),\label{eq:initialPointObj}\\
\st\  & \delta_{k}+\frac{1}{\Gamma_{k}}\bar{f}_{k}\big(\mathbf{x}_{k},\boldsymbol{\theta};\mathbf{x}_{k}^{(i)},\boldsymbol{\theta}^{(i)}\big)\nonumber \\
 & {\small \quad\geq\sigma_{k}^{2}+ \sum \nolimits_{\ell\in\mathcal{M}\setminus\{k\}}\!\!\!\!\big(\wp_{k\ell}^{2}+\bar{\wp}_{k\ell}^{2}\big)\ \forall k\in\mathcal{K},\label{eq:intialC1}}\\
\  & \bar{\delta}_{k}+\sigma_{\mathrm{T}}^{2}+\sum_{\ell\in\mathcal{M}\setminus\{k\}}f_{\ell}\big(\tilde{\mathbf{x}}_{\ell},\boldsymbol{\theta};\tilde{\mathbf{x}}_{\ell}^{(i)},\boldsymbol{\theta}^{(i)}\big)\nonumber \\
 & \qquad\geq\frac{1}{\hat{\Gamma}_{k}}\big(\tau_{k}^{2}+\bar{\tau}_{k}^{2}\big)\ \forall k\in\mathcal{K},\label{eq:initialC2}\\
 & \eqref{eq:TPC},\eqref{eq:UMC},\eqref{eq:etaChi},\eqref{eq:transSINRC-2},\nonumber \\
 & \delta_{k}\geq0,\bar{\delta}_{k}\geq0\ \forall k\in\mathcal{K}.\label{eq:initialC3}
\end{align}
\end{subequations}
Note that the problem in~\eqref{eq:initialPointProb} is always feasible
for sufficiently large $\boldsymbol{\delta}$ and $\bar{\boldsymbol{\delta}}$.
We solve the problem in~\eqref{eq:initialPointProb} by following
a similar procedure to that of~\textbf{Algorithm~\ref{algo}}, with
random $\mathbf{X}$ and $\boldsymbol{\theta}$ as initial points.
The minimization in \ref{eq:initialPointProb} forces $\delta_{k}$
and $\bar{\delta}_{k}$ to approach $0$. At convergence, if $\delta_{k}=\bar{\delta}_{k}=0\ \forall k\in\mathcal{K}$,
the problem in~\eqref{eq:transProb-1} is obviously feasible. Thus
we can choose the final values of $\mathbf{X}$ and $\boldsymbol{\theta}$
in~\eqref{eq:initialPointProb} as initial points for~\textbf{Algorithm~\ref{algo}}.
However, if the objective $\sum_{k\in\mathcal{K}}(\delta_{k}+\bar{\delta}_{k})$
is not zero at convergence, then we simply declare that the considered
problem is infeasible and will not run \textbf{Algorithm~\ref{algo}}\footnote{We note that the considered problem may be feasible even though $\sum_{k\in\mathcal{K}}(\delta_{k}+\bar{\delta}_{k})>0$.
The reason is that the SCA-based method applied to solve \eqref{eq:initialPointProb}
can only guarantee a stationary solution. In general, checking \eqref{eq:mainProb}
is feasible or not is an NP hard problem since the feasible set is
non-convex. For practical purposes, if the SCA-based method cannot
find a feasible solution, we can simply say that the problem is infeasible
and ignore this realization.} 
\end{rem}
The convergence of the proposed SCA-based method in~\textbf{Algorithm~\ref{algo}
}can be readily proved following the set of arguments in~\cite[Sec.~III-A]{23-WCL-SCA}.

\subsection{Complexity Analysis}

It is straightforward to show that the total number of (real-valued)
optimization variables in~\eqref{eq:transProb-1} is $2\big(L^{2}+K^{2}+2KL+N\big)+1$,
and the total number of (second-order) conic constraints is $4K^{2}+4KL+2K+N+2$.
Therefore, following the arguments in~\cite[Sec.~6.6.2]{ModernOptLectures},
the overall per-iteration complexity of the proposed SOCP-based method
is given by 
\begin{align}
 & \mathcal{O}\big[\big(4K^{2}+4KL+N\big)^{0.5}\big(2K^{2}+4KL+2L^{2}+2N)\nonumber \\
 & \big(4K^{5}+8K^{4}L+4K^{3}L^{2}+48K^{3}L+60K^{2}L^{2}+24KL^{3}\nonumber \\
 & +52KL^{2}+4L^{4}+\big(2K^{2}+4KL+2L^{2}+2N\big)^{2}\big)\big].\label{eq:exactComplexity}
\end{align}
However, in a practical setup, the number of elements in the IRS is
expected to be much larger than the number of BS antennas and the
number of users, i.e., $N\gg\max\{L,K\}$. Hence, the complexity of
the proposed SCA-based method can be well-approximated by $\mathcal{O}\big(N^{3.5}\big)$.
On the other hand, the per-iteration computational complexity of~\cite[Algorithm~1]{22-arXiv-sISAC}
can be approximated by $\mathcal{O}(N^{3})$ (see~\cite[Sec.~III-C]{22-arXiv-sISAC}).
Although the order of complexity of the proposed SCA-based method
is slightly higher than that of the penalty-based benchmark, we will
show in the simulation section that the proposed method requires fewer
iterations, resulting in a significantly reduced problem-solving time.
\begin{figure*}[t]
	\begin{minipage}[t]{0.32\textwidth}%
		\centering
		\input{fig_convSeq.tex}\vspace{-0.2cm}
		\caption{Convergence results for $L=4$, $K=3$, $N=100$, $P=40$~dBm, $\Gamma_{k}=10$~dB $\forall k\in\mathcal{K}$ and $\hat{\Gamma}_{k}=0$~dB $\forall k\in\mathcal{K}$.}
		\label{fig:conv}%
	\end{minipage}
	\hfill
	\begin{minipage}[t]{0.32\textwidth}%
		\centering
		\input{fig_BPG.tex}\vspace{-0.2cm}
		\caption{Average beampattern gain for $L=4$, $K=3$,$\Gamma_{k}=10$~dB $\forall k\in\mathcal{K}$ and $\hat{\Gamma}_{k}=0$~dB $\forall k\in\mathcal{K}$.}
		\label{fig:BPG}
	\end{minipage}
	\hfill
	\begin{minipage}[t]{0.32\textwidth}%
		\centering
		\input{fig_runTime.tex}\vspace{-0.2cm}
		\caption{Average problem solving time for $L=4$, $P=40$~dBm$,\Gamma_{k}=10$~dB $\forall k\in\mathcal{K}$ and $\hat{\Gamma}_{k}=0$~dB $\forall k\in\mathcal{K}$.}
		\label{fig:runTime}%
	\end{minipage}
\end{figure*}

\section{\label{sec:Results}Numerical Results and Discussion}

In this section, we present a detailed performance comparison between
the proposed SCA-based method and the penalty-based benchmark approach
of~\cite[Algorithm~1]{22-arXiv-sISAC}. The location of the nodes
and the channel model assumed here are the same as those in~\cite{22-arXiv-sISAC}.
The simulations are performed on a high-performance computing cluster
with a Intel Xeon Gold 6152 processor, using Python v3.9.7 and MOSEK
Fusion API for Python Rel.-10.0.40~\cite{mosek}. In Figs.~\ref{fig:BPG}
and~\ref{fig:runTime}, the results are obtained by averaging over
100 independent channel realizations.

In Fig.~\ref{fig:conv}, we show the convergence behavior of both
the proposed and penalty-based benchmark methods. For the given set
of channels, the proposed SCA-based method converges in less than
30 iterations, whereas the penalty-based benchmark requires around
270 iterations. Nevertheless, even with significantly fewer iterations,
the proposed method results in nearly a 30\% higher beampattern gain
as compared to that offered by the penalty-based benchmark. More interestingly,
each iteration of the proposed SCA-based method returns a set of feasible
points and therefore the iterations of the proposed method can be
terminated even before convergence has been attained, if this is required.
On the other hand, the penalty-based benchmark returns a feasible
solution only in the final outer-loop iteration, and therefore the
algorithm cannot be stopped earlier to achieve a feasible solution.
Therefore, the benchmark is not suitable in rapidly changing environments
with very small coherence times where at least a suboptimal solution
is required within a certain fraction of the channel's coherence time.

The impact of the number of IRS elements on the average beampattern
gain for the two algorithms- is shown in Fig.~\ref{fig:BPG}. An
increase in the number of IRS elements increases the degrees-of-freedom
at the IRS, allowing the IRS to perform highly focused beamforming.
This in turn results in increasing beampattern gain with increasing
$N$. On the other hand, since a fixed amount of transmit power is
required to achieve the SINR constraints at the communication users,
a higher transmit power budget results in a higher surplus power at
the BS, which is then used to attain a larger beampattern gain toward
the target. Therefore, increasing the value of $P$ increases the
average beampattern gain, which is also clearly evident from the figure.
The performance gap between the SCA-based and penalty-based methods
increases with an increase in the number of IRS elements. As the value
of $N$ increases, the impact of coupling between $\mathbf{X}$ and
$\boldsymbol{\theta}$ becomes more intricate. Therefore, the solution
obtained via the AO-based approach of~\cite{22-arXiv-sISAC} returns
a highly suboptimal beampattern gain. On the other hand, as clearly
observed in the figure, the simultaneous update of all variables in
the proposed algorithm outperforms the penalty-based benchmark. 

In Fig.~\ref{fig:runTime}, we plot the average problem-solving time
versus the number of IRS elements for different numbers of communication
users $K$. As the value of $N$ and/or $K$ increases, the size of
the optimization problem to be solved also increases for both methods.
This in turn increases the average problem solving time for both approaches.
Although the per-iteration complexity of the proposed SCA-based method
is slightly higher than that of the penalty-based benchmark, the proposed
approach requires a much smaller time to find the solution due to
its convergence in fewer iterations.

\section{Conclusion}

In this paper, we have considered the problem of optimal transmit
and reflective beamforming design in a secure IRS-enabled ISAC system.
More specifically, we aim to maximize the beampattern gain toward
the eavesdropping target subject to the SINR constraints at the communication
users and information leakage constraints at the target. In contrast
to the conventional AO-based approach, we proposed a novel SCA-based
optimization in which all variables are updated simultaneously. The
superiority of the proposed method was clearly established with the
help of numerical experiments in terms of both achieving a high-performance
solution and low problem-solving time compared to that of the penalty-based
benchmark. Moreover, the performance gap between the proposed SCA-based
approach and penalty-based benchmark was shown to be increasing with
the number of IRS elements or the transmit power budget.
\balance
\bibliographystyle{IEEEtran}
\bibliography{v2_Vaibhav}

\end{document}

%% file: fig_convSeq.tex
\newcommand{\vasymptote}[2][]{
    \draw [densely dashed,#1] ({rel axis cs:0,0} -| {axis cs:#2,0}) -- ({rel axis cs:0,1} -| {axis cs:#2,0});
}
\resizebox{0.95\columnwidth}{0.8\columnwidth}{
\begin{tikzpicture}[/pgfplots/tick scale binop=\times]
  \begin{axis}[
    xlabel={Iteration number},
	xtick={1,50,100,150,200,250,300},
	xmin=0,xmax=330,
	ylabel={Instantaneous beampattern gain},
	ylabel near ticks,
	ymin=-7e-7,ymax=7.5e-6,
	grid=both,
	minor grid style={gray!25},
	major grid style={gray!25},
	legend columns=1, 
legend style={{nodes={scale=0.9, transform shape}}, at={(0.213,0)},  anchor=south west, draw=black,fill=white,legend cell align=left,inner sep=1pt,row sep = -2pt}]
	\addplot[line width=1pt,solid,color=blue] table [y=proposed, x=iter,col sep = comma]{convComp.csv};
	\addlegendentry{\small{Proposed (true objective)}}
	\addplot[line width=1pt,dashdotted,color=myred] table [y=augBPG, x=iter,col sep = comma]{convComp.csv};
	\addlegendentry{\small{Benchmark~[14] (augmented objective)}}
	\addplot[line width=1pt,densely dashed,color=mygreen] table [y=trueBPG, x=iter,col sep = comma]{convComp.csv};
	\addlegendentry{\small{Benchmark~[14] (true objective)}}
	\end{axis}
\end{tikzpicture}
}

%% file: fig_BPG.tex
\newcommand{\vasymptote}[2][]{
    \draw [densely dashed,#1] ({rel axis cs:0,0} -| {axis cs:#2,0}) -- ({rel axis cs:0,1} -| {axis cs:#2,0});
}
\resizebox{0.95\columnwidth}{0.8\columnwidth}{
\begin{tikzpicture}[/pgfplots/tick scale binop=\times]
  \begin{axis}[
    xlabel={$N$},
	xtick={64,144,256,400,576},
	xticklabels={$8^2$,$12^2$,$16^2$,$20^2$,$24^2$},
	xmin=64,xmax=576,
	ylabel={Average beampattern gain},
	ylabel near ticks,
	ymin=0,ymax=4.1e-4,
	grid=both,
	minor grid style={draw=none},
	major grid style={gray!25},
	legend columns=1, 
legend style={{nodes={scale=0.9, transform shape}}, at={(0,0.636)},  anchor=south west, draw=black,fill=white,legend cell align=left,inner sep=1pt,row sep = -2pt}]
	\addplot[line width=1pt,solid,color=mygreen,mark=*,mark options={solid,mygreen},mark size=1.5pt] table [y=proposed45dBm,x=Ns,col sep = comma]{bpg_vs_Ns.csv};
	\addlegendentry{Proposed: $P = 45$~dBm}
		
	\addplot[line width=1pt,solid,color=myred,mark=square*,mark options={solid,myred},mark size=1.5pt] table [y=proposed40dBm,x=Ns,col sep = comma]{bpg_vs_Ns.csv};
	\addlegendentry{Proposed: $P = 40$~dBm}

	\addplot[line width=1pt,solid,color=blue,mark=triangle*,mark options={solid,blue},mark size=2.5pt] table [y=proposed35dBm,x=Ns,col sep = comma]{bpg_vs_Ns.csv};
	\addlegendentry{Proposed: $P = 35$~dBm}
	
	\addplot[line width=1pt,densely dashed,color=mygreen,mark=o,mark options={solid,mygreen},mark size=1.5pt] table [y=benchmark45dBm,x=Ns,col sep = comma]{bpg_vs_Ns.csv};
	\addlegendentry{Benchmark~[14]: $P = 45$~dBm}
	
	\addplot[line width=1pt,densely dashed,color=myred,mark=square,mark options={solid,myred},mark size=1.5pt] table [y=benchmark40dBm,x=Ns,col sep = comma]{bpg_vs_Ns.csv};
	\addlegendentry{Benchmark~[14]: $P = 40$~dBm}
	
	\addplot[line width=1pt,densely dashed,color=blue,mark=triangle,mark options={solid,blue},mark size=2.5pt] table [y=benchmark35dBm,x=Ns,col sep = comma]{bpg_vs_Ns.csv};
	\addlegendentry{Benchmark~[14]: $P = 35$~dBm}

	\end{axis}
\end{tikzpicture}
}

%% file: fig_runTime.tex
\newcommand{\vasymptote}[2][]{
    \draw [densely dashed,#1] ({rel axis cs:0,0} -| {axis cs:#2,0}) -- ({rel axis cs:0,1} -| {axis cs:#2,0});
}
\resizebox{0.95\columnwidth}{0.8\columnwidth}{
\begin{tikzpicture}
  \begin{axis}[
	log basis y={10},
    xlabel={$N$},
	xtick={64,144,256,400,576},
	xticklabels={$8^2$,$12^2$,$16^2$,$20^2$,$24^2$},
	xmin=64,xmax=576,
	ylabel={Average problem-solving time (s)},
	ylabel near ticks,
	ytick={0,500,1000,1500,2200},
	ymin=1,ymax=2200,
	grid=both,
	minor grid style={draw=none},
	major grid style={gray!25},
	legend columns=1, 
legend style={{nodes={scale=0.9, transform shape}}, at={(0,0.635)},  anchor=south west, draw=black,fill=white,legend cell align=left,inner sep=1pt,row sep = -2pt}]
	\addplot[line width=1pt,densely dashed,color=mygreen,mark=o,mark options={solid,mygreen},mark size=1.5pt] table [y=benchmark4U, x=Ns,col sep = comma]{time_vs_Ns.csv};
	\addlegendentry{Benchmark~[14]: $K = 4$}
		
	\addplot[line width=1pt,densely dashed,color=myred,mark=square,mark options={solid,myred},mark size=1.5pt] table [y=benchmark3U, x=Ns,col sep = comma]{time_vs_Ns.csv};
	\addlegendentry{Benchmark~[14]: $K = 3$}

	\addplot[line width=1pt,densely dashed,color=blue,mark=triangle,mark options={solid,blue},mark size=2pt] table [y=benchmark2U, x=Ns,col sep = comma]{time_vs_Ns.csv};
	\addlegendentry{Benchmark~[14]: $K = 2$}

	\addplot[line width=1pt,solid,color=mygreen,mark=*,mark options={solid,mygreen},mark size=1.5pt] table [y=proposed4U, x=Ns,col sep = comma]{time_vs_Ns.csv};
	\addlegendentry{Proposed: $K = 4$}

	\addplot[line width=1pt,solid,color=myred,mark=square*,mark options={solid,myred},mark size=1.5pt] table [y=proposed3U, x=Ns,col sep = comma]{time_vs_Ns.csv};
	\addlegendentry{Proposed: $K = 3$}
		
	\addplot[line width=1pt,solid,color=blue,mark=triangle*,mark options={solid,blue},mark size=2pt] table [y=proposed2U, x=Ns,col sep = comma]{time_vs_Ns.csv};
	\addlegendentry{Proposed: $K = 2$}

	\end{axis}
\end{tikzpicture}
}